\pdfoutput=1  %arXiv addition for pdflatex, permits no file extensions.
\documentclass[]{article}  
\usepackage{url,float}
\usepackage{graphicx}
\usepackage{amsfonts}
\usepackage{amssymb}
\usepackage{latexsym}

\newcommand{\hide}[1]{}

\newcommand{\ABox}{
\raisebox{3pt}{\framebox[6pt]{\rule{6pt}{0pt}}}
}
\newenvironment{proof}{{\bf Proof:}}{\hfill\ABox}

\newtheorem{theorem}{{\bf Theorem}}

\newtheorem{lemma}[theorem]{Lemma}

\newcommand{\tablab}[1]{\label{tab:#1}}
\newcommand{\figlab}[1]{\label{fig:#1}}

\newcommand{\eqref}[1]{\ref{eq:#1}}
\newcommand{\figref}[1]{\ref{fig:#1}}
\newcommand{\tabref}[1]{\ref{tab:#1}}

{\makeatletter
 \gdef\xxxmark{%
   \expandafter\ifx\csname @mpargs\endcsname\relax % in minipage?
     \expandafter\ifx\csname @captype\endcsname\relax % in figure/caption?
       \marginpar{xxx}% not in a caption or minipage, can use marginpar
     \else
       xxx % notice trailing space
     \fi
   \else
     xxx % notice trailing space
   \fi}
 \gdef\xxx{\@ifnextchar[\xxx@lab\xxx@nolab}
 \long\gdef\xxx@lab[#1]#2{{\bf [\xxxmark #2 ---{\sc #1}]}}
 \long\gdef\xxx@nolab#1{{\bf [\xxxmark #1]}}
 \gdef\turnoffxxx{\long\gdef\xxx@lab[##1]##2{}\long\gdef\xxx@nolab##1{}}%
}

\title{Some Properties of Yao $Y_4$ Subgraphs}
\author{%
Joseph O'Rourke%
    \thanks{Department of Computer Science, Smith College, Northampton, MA
      01063, USA.
      \protect\url{orourke@cs.smith.edu}.}
}%author

\begin{document}
\maketitle

\begin{abstract}
The Yao graph for $k=4$, $Y_4$, 
is naturally partitioned into four subgraphs, one per quadrant.
We show that the subgraphs for one quadrant differ from the subgraphs for
two adjacent quadrants in three properties:
planarity, connectedness, and whether the directed graphs are spanners.
\end{abstract}

\section{Introduction}
The Yao graph is defined for an integer parameter $k$;
here we study only $k=4$, and call
$\overrightarrow{Y_4}$ the directed Yao graph, and $Y_4$ the undirected version.
For a set of points $P$, $\overrightarrow{Y_4}$ connects each point
to its closest neighbor in each of the four quadrants surrounding it,
defined as in
%The quadrant definition is illustrated in 
Figure~\figref{QuadrantsDefinition}.
Ties are broken arbitrarily.
%%%%%%%%%%%%%%%%%%%%%%%%%%%%%%%%%Figure Begin
\begin{figure}[htbp]
\centering
\includegraphics[width=0.5\linewidth]{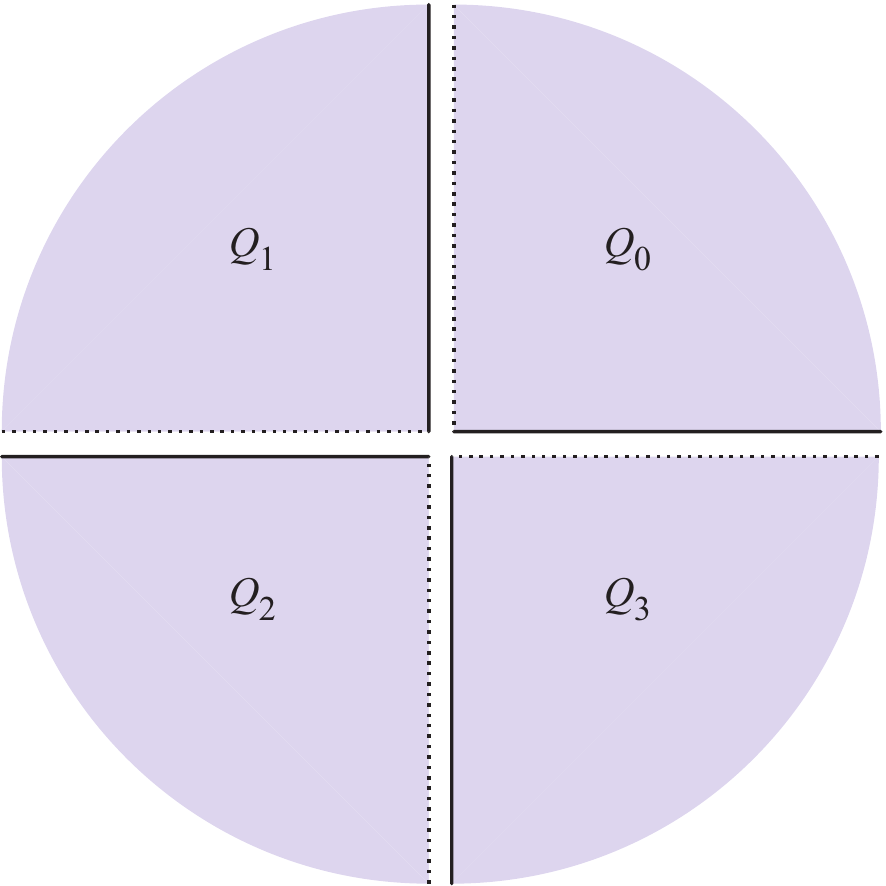}
\caption{Definition of quadrants. Solid lines are closed, dotted lines are open.}
\figlab{QuadrantsDefinition}
\end{figure}
%%%%%%%%%%%%%%%%%%%%%%%%%%%%%%%%%Figure End
The undirected graph $Y_4$ simply ignores the direction.

The question of whether $Y_4$ is a spanner was raised in
\cite{sppyg-dmp-09}.
A $t$-spanner has the property that the path between $a$ and $b$ in the
graph is no longer than $t |a b|$, for a constant $t$.
In this note, we do not further motivate the study of $Y_4$,
but rather investigate some properties of subgraphs of $Y_4$, which
may ultimately have some bearing on whether it is a spanner.

We make two ``general position'' assumptions:
\begin{enumerate}
\item No two pair of points determine the same distance
(so there are no ties).
\item
No two points share a vertical or horizontal coordinate.
\end{enumerate}
These assumptions simplify the presentation.
In this note, we will not explore whether the assumptions can be removed while retaining
all the results.

\paragraph{Notation.}
$Q_i(a,b)$ is the circular quadrant whose origin is at $a$ and which reaches out to $b$.
Often the subscript $i$ will be dropped, as it is determined by $a$ and $b$.
$Q_i(a)$ is the unbounded quadrant with corner at $a$.
Thus,
$Q_i(a,b) = Q_i(a) \cap \mathrm{disk}(a, |ab|)$.
$R(a,b)$ is the closed rectangle with opposite corners $a$ and $b$.
%This is an easy but important fact:
%$$R(a,b) \subset Q(a,b) \;.$$
%In particular, when $ab \in Y_4$, then it must be that $Q(a,b)$ is empty of points,
%and therefore so must $R(a,b)$ be empty.
%(Note that the unique-distance assumption guarantees that the closed
%$Q(a,b)$ is empty of points.)

We focus on two adjacent quadrants, $Q_0$ and $Q_1$.
Let $Y_4^{\{\lambda\}}$ be the $Y_4$ graph restricted to the 
quadrants in the list $\lambda$.
See Figure~\figref{Y40} for examples.
%%%%%%%%%%%%%%%%%%%%%%%%%%%%%%%%%Figure Begin
\begin{figure}[htbp]
\centering
\includegraphics[height=0.9\textheight]{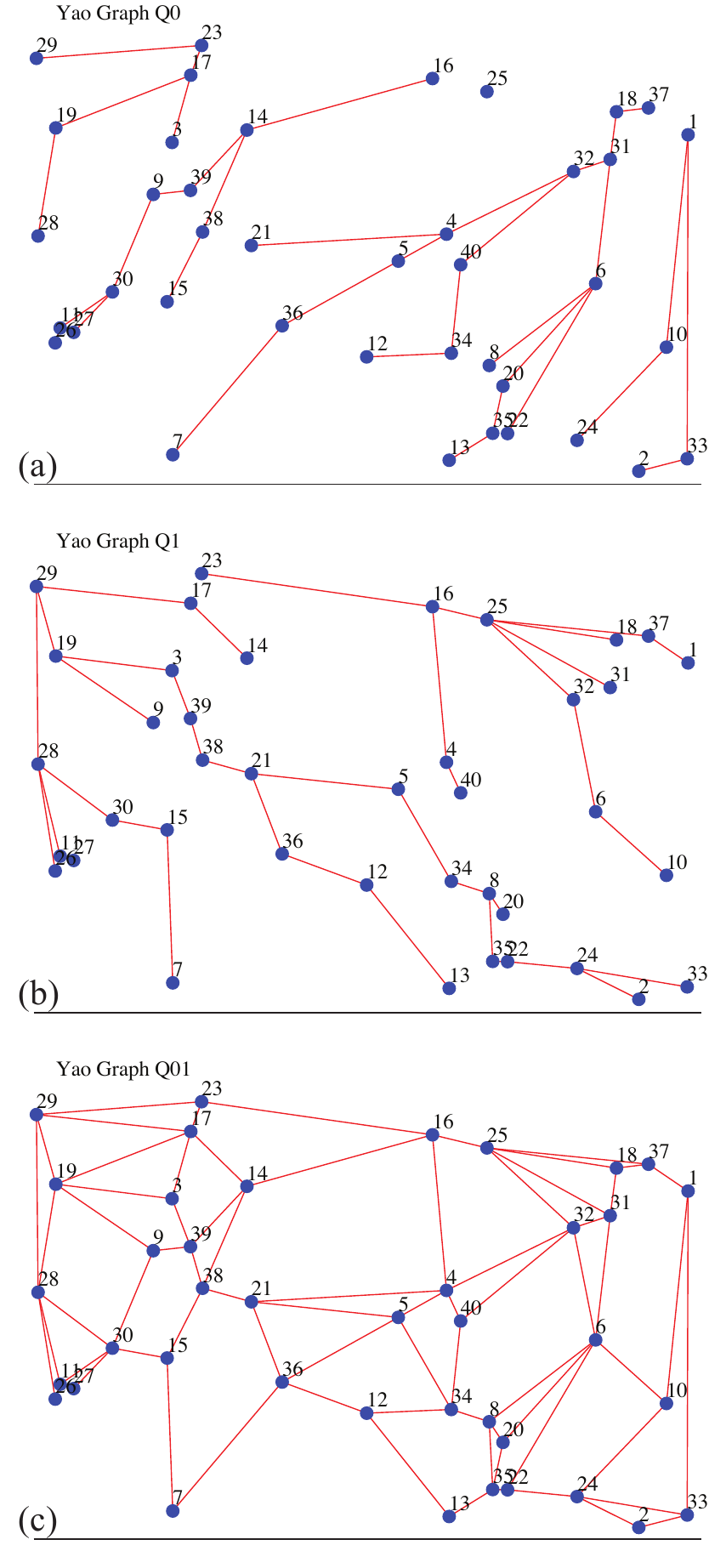}
\caption{$Y_4^{\{0\}}$, $Y_4^{\{1\}}$, and $Y_4^{\{0,1\}}$, for the same $40$-point set.}
\figlab{Y40}
\end{figure}
%%%%%%%%%%%%%%%%%%%%%%%%%%%%%%%%%Figure End

Our results are summarized in Table~\tabref{Results}.
\begin{table}[htbp]
\begin{center}
\begin{tabular}{| l || c | c |}
\hline
\emph{Property} & $Y_4^{\{i\}}$ & $Y_4^{\{i,i+1\}}$ 
\\ \hline \hline
Planarity & planar & not planar
\\ \hline
Connectedness & not connected & connected 
\\ \hline
Undirected spanner & not a spanner & not a spanner 
\\ \hline
Directed spanner & spanner & not a spanner
\\ \hline
\end{tabular}
\end{center}
\tablab{Results}
\caption{Summary of Results}
\end{table}

\section{Planarity}
It is known that $Y_4^{\{i\}}$ is a planar forest, in general disconnected;
see Figure~\figref{Y40}(a,b).
This is folklore,\footnote{
Mirela Damian [private communication, Feb. 2009].
}
but we offer a proof of planarity.

\begin{lemma}
No two edges of $Y_4^{\{i\}}$ properly cross.
\end{lemma}
\begin{proof}
Let both $ab$ and $cd$ be in $Y_4^{\{0\}}$, and suppose $ab$ and $cd$ properly cross.
see Figure~\figref{Q0noncrossing}.
%%%%%%%%%%%%%%%%%%%%%%%%%%%%%%%%%Figure Begin
\begin{figure}[htbp]
\centering
\includegraphics[width=\linewidth]{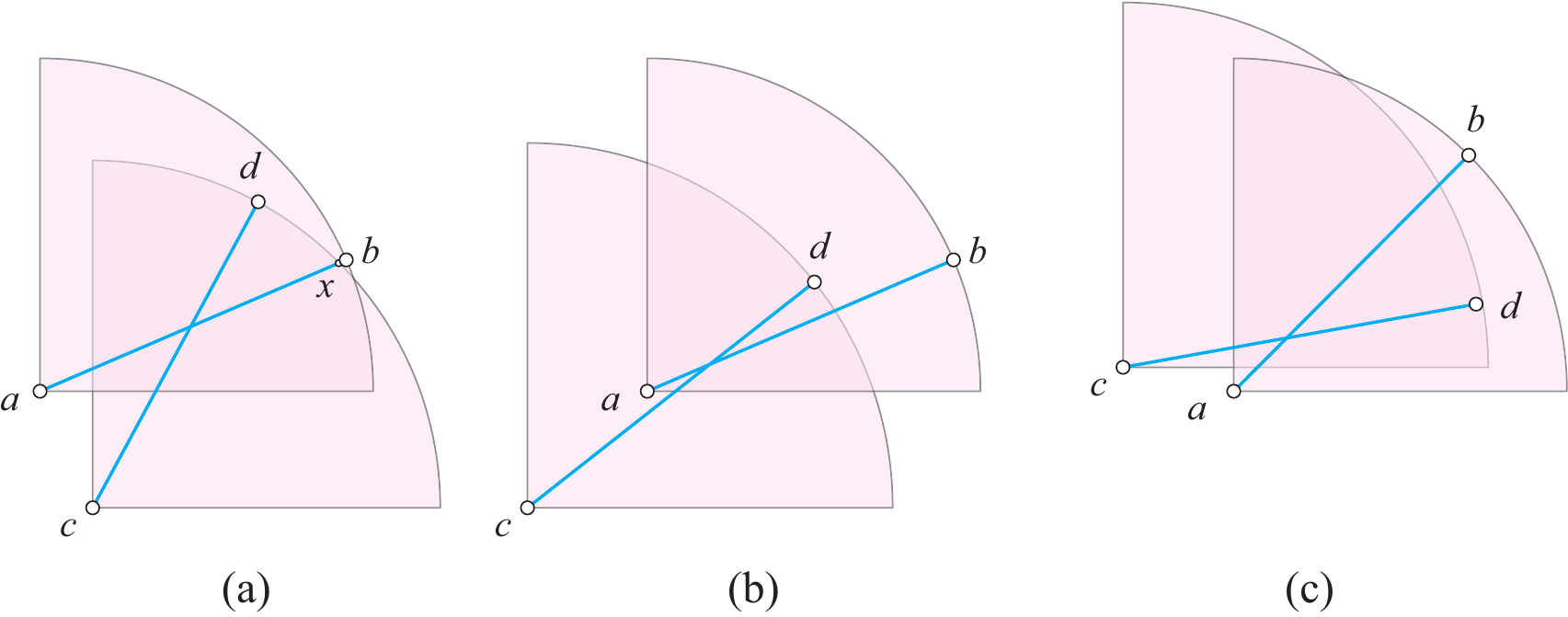}
\caption{$ab$ and $cd$ may not cross.}
\figlab{Q0noncrossing}
\end{figure}
%%%%%%%%%%%%%%%%%%%%%%%%%%%%%%%%%Figure End
The quadrants $Q(a,b)$ and $Q(c,d)$ must be empty of points.
We consider three cases, depending on the location of $c$ w.r.t. $a$.
\begin{enumerate}
\item $c \in Q_3(a)$.
Then $cd$ crosses $ab$ from below.  We analyze just this case in detail.
Because $b \not\in Q(c,d)$, the circular boundary of $Q(c,d)$ must cut $ab$,
say at $x$.  Consider two further cases
\begin{enumerate}
\item The slope of the arc of $Q(c,d)$ at $x$ is shallower than the slope of the
arc of $Q(a,b)$ at $b$; see Figure~\figref{Q0noncrossing}(a).
Then $d \in Q(a,b)$.
\item The slope at $x$ is equal to or steeper than that at $b$.
Then, because $c$ is strictly below $a$, the radius $|cd|$ is greater than
$|ab|$.
But then $c$ cannot be in $Q_3(a)$.
\end{enumerate}
\item $c \in Q_2(a)$.
Then $cd$ could cross $ab$ from below, Figure~\figref{Q0noncrossing}(b),
or from above, Figure~\figref{Q0noncrossing}(c).
In both cases, a quadrant that must be empty is not.
\item $c \in Q_1(a)$.
This case is the same as the first case, with the roles of $a$ and $c$ interchanged.
\end{enumerate}
\end{proof}

In contrast, $Y_4^{\{i,i+1\}}$ may be nonplanar.
Figure~\figref{NonPlanar01}(a) shows two crossing edges;
(b) shows the full graph $Y_4^{\{0,1\}}$.
%%%%%%%%%%%%%%%%%%%%%%%%%%%%%%%%%Figure Begin
\begin{figure}[htbp]
\centering
\includegraphics[width=0.75\linewidth]{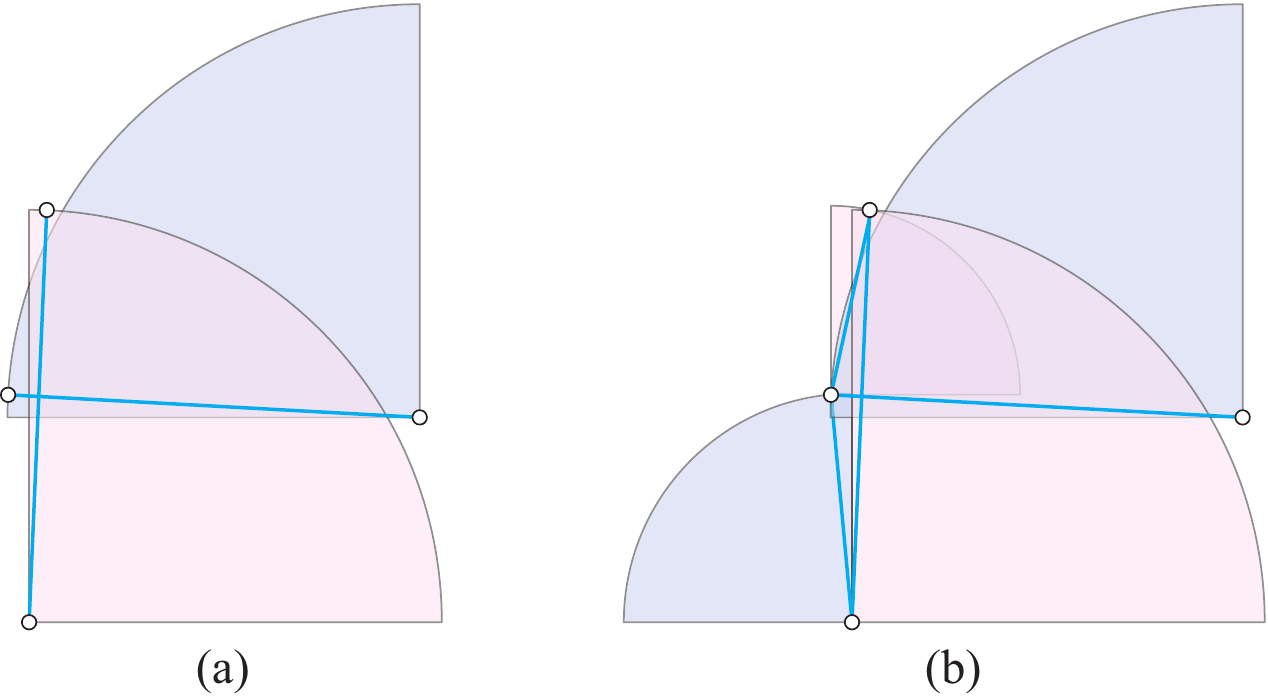}
\caption{$Y_4^{\{0,1\}}$ can be nonplanar.}
\figlab{NonPlanar01}
\end{figure}
%%%%%%%%%%%%%%%%%%%%%%%%%%%%%%%%%Figure End

As should be evident from Figure~\figref{Y40}(c),
crossing edges are rare, requiring precise placement of four points.
Although it would be difficult to quantify, a ``typical'' $Y_4^{\{i,i+1\}}$ graph
is planar.

\section{Connectedness}
We can see in Figure~\figref{Y40}(a,b)
that
$Y_4^{\{i\}}$ is, in general, disconnected.
In contrast, $Y_4^{\{i,i+1\}}$ is connected.
See again Figure~\figref{Y40}(c).

\begin{lemma}
$\overrightarrow{Y_4^{\{i,i+1\}}}$ is a connected graph.
\end{lemma}
\begin{proof}
We choose $i{=}0$ w.l.o.g.
So we are concerned with upward $+y$-connections, in $Q_0$ and $Q_1$.
The proof is by induction on the number of points $n$ in the set $P$.
The basis of the induction is trivial, for an $n{=}1$ point set is connected.
Let $P$ have $n>1$ points, and let $a$ be the point with the lowest
$y$-coordinate.  
By Assumption~(2), $a$ is unique.  

Delete this from $P$, reducing to a point set $P'$ with $|P'| = n{-}1$.  
Then the set of points $P' = P \setminus \{a\}$
satisfies the induction hypothesis, and so is connected into a graph $\overrightarrow{G'}$. 
See Figure~\figref{Connected01a}.
Put back point $a$.
Because all the quadrants determining edges $\overrightarrow{bc} \in \overrightarrow{G'}$ are $Q_0$ or $Q_1$, they lie at or above $b_y$, the $y$-coordinate of the lowest point
in $P'$,
$b$.
Thus $a$ cannot lie in any quadrant, and so adding $a$ to $P'$ does not break any edge
of $\overrightarrow{G'}$.\footnote{
    Note that if the induction instead removed the topmost point from $P$, this claim
    would no longer hold.}
Finally, $a$ itself must have at least one outgoing edge upward, for $Q_0$ and $Q_1$ cover
the half-plane above $a_y$, which contains at least one point of $P'$.
\end{proof}

%%%%%%%%%%%%%%%%%%%%%%%%%%%%%%%%%Figure Begin
\begin{figure}[htbp]
\centering
\includegraphics[width=0.5\linewidth]{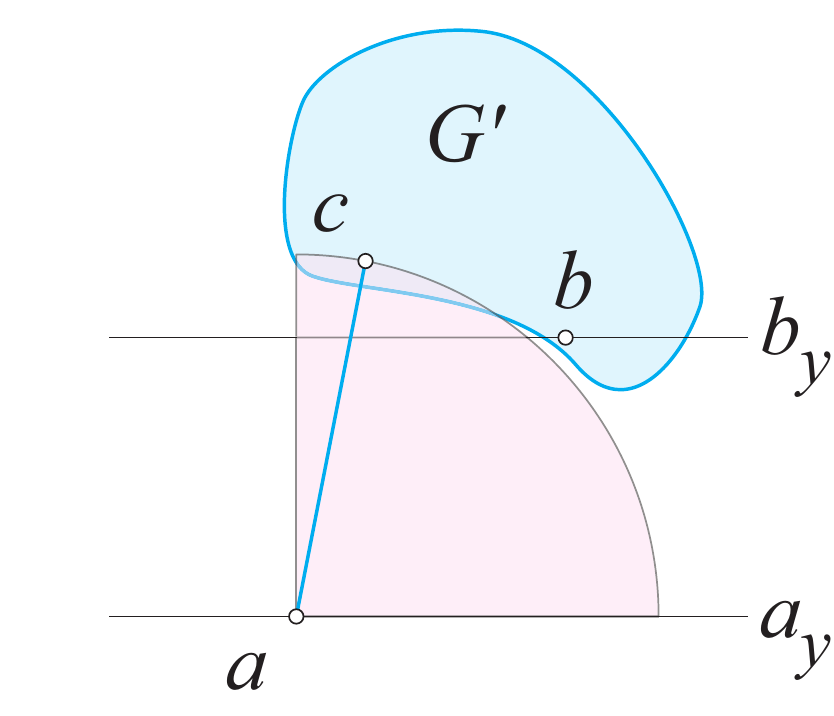}
\caption{$Y_4^{0,1}$ must be connected.}
\figlab{Connected{01}a}
\end{figure}
%%%%%%%%%%%%%%%%%%%%%%%%%%%%%%%%%Figure End

\section{Undirected Spanners}
It is clear that $Y_4^{\{i\}}$ is not a spanner, because it may be
disconnected.  Points on a negatively sloped line result in a completely
disconnected graph of isolated points.
Neither is $Y_4^{\{i,i+1\}}$ a spanner.  Points uniformly spaced on two lines forming a `$\Lambda$' shape
%(upsidedown `V') 
both have directed paths up to the apex in $Y_4^{\{0,1\}}$, 
but the leftmost and rightmost lowest
points can be arbitrarily far apart in the graph.

\section{Directed Spanners}
We turn then to directed versions of these questions.
Call a directed graph a \emph{directed spanner} if every directed path
is no more than $t$ times the path's end-to-end Euclidean distance,
for $t$ a constant.

\begin{lemma}
$\overrightarrow{Y_4^{\{i\}}}$ is a directed spanner:
no directed path is more than $\sqrt{2}$ times the end-to-end Euclidean distance.
\end{lemma}
\begin{proof}
Let $a$ and $b$ be the endpoints of the path.
Then the path is an $xy$-monotone path remaining inside
$R(a,b)$.  Therefore its length is at most half the perimeter of this
rectangle, which is at most $\sqrt{2}$ times the diagonal length.
\end{proof}

\begin{lemma}
$\overrightarrow{Y_4^{\{i,i+1\}}}$ is not a directed spanner:
directed paths can be arbitrarily long:
more than any constant $t>1$ times the end-to-end Euclidean distance.
\end{lemma}
\begin{proof}
Consider the path $(a,b,c,d)$ in Figure~\figref{Q01longpath}(a).
%%%%%%%%%%%%%%%%%%%%%%%%%%%%%%%%%Figure Begin
\begin{figure}[htbp]
\centering
\includegraphics[width=0.99\linewidth]{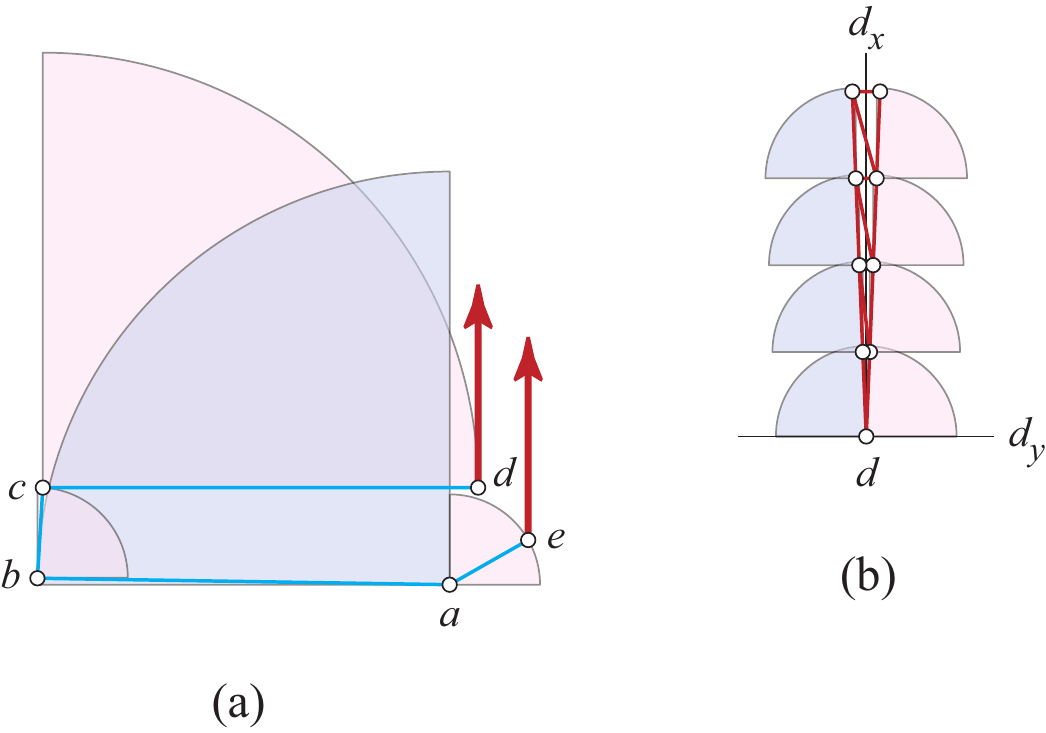}
\caption{An arbitrarily long path in $Y_4^{\{0,1\}}$ .}
\figlab{Q01longpath}
\end{figure}
%%%%%%%%%%%%%%%%%%%%%%%%%%%%%%%%%Figure End
It is clear that this path can be made arbitrarily long with respect to $|ad|$,
by lowering the vertical coordinates of $c$ and $d$.
Now we show how to avoid any other directed connection between $a$ and $d$.

Let the other outgoing edge from $a$ go to $e$ as shown.
We now direct paths from $d$ and from $e$ that do not connect.
The idea is depicted in Figure~\figref{Q01longpath}(b).
We create a series of nearly vertical paths from $d$, and from $e$.
Above $d=(d_x,d_y)$, two points are placed at $(d_x \pm \epsilon, d_y+1)$,
$0 < \epsilon \ll 1$.
The two outgoing edges from $d$ will terminate on these.
Then above those we place two more points at $(d_x \pm 2 \epsilon, d_y+2)$.
Now we get both upward and diagonal connections among the four points,
with one ``diagonal'' being horizontal.\footnote{
	The definition in Figure~\protect\figref{QuadrantsDefinition} shows that
	 $(d_x - \epsilon, d_y+1)$ will connect horizontally to $(d_x + \epsilon, d_y+1)$.
	 }
The point is that all the outgoing edges are
accounted for.

Repeating this construction, we can make a nearly vertical tower of points,
connected by vertical paths, but otherwise insulated from one another.
So the only path from $a$ to $d$ is $(a,b,c,d)$.
\end{proof}

%\section{Removing General Position Assumptions}
\section{Future Work}
The obvious next step is to examine properties of three
quadrants, $Y_4^{\{i,i+1,i+2\}}$, before finally tackling $Y_4$ itself.

%%%%%%%%%%%%%%%%%%%%%%%%%%%%%%%%%%%%%%%%%%%%%%%%%%%%%%%%%%%%%%%%%%%%%%%%%%%%
%\small
%\baselineskip=0.9\baselineskip
% Decrease the space between bibliography items.
%\let\realbibitem=\bibitem
%\def\bibitem{\par \vspace{-0.6ex}\realbibitem}

\bibliographystyle{alpha}
\bibliography{/home/orourke/bib/geom/geom}
\end{document}